\documentclass[12pt]{article}
\usepackage{amsmath}
\usepackage{amssymb}
\usepackage{epsfig}
\newcommand{\om}{\omega}
\newcommand{\as}{\alpha_\mathrm{s}}
\newcommand{\asb}{{\bar \alpha}_\mathrm{s}}
\newcommand{\tbar}{{\bar t}}
\newcommand{\fbar}{{\bar f}}
\newcommand{\gambar}{\bar \gamma}
\newcommand{\order}[1]{{\cal O}\left(#1\right)}
\newcommand{\ie}{\textrm{i.e.}\ }
\newcommand{\CF}{{\cal F}}
\newcommand{\oms}{\omega_\mathrm{s}}
\newcommand{\omp}{\omega_\mathrm{I\!P}}
\newcommand{\pom}{\mathrm{I\!P}}
\newcommand{\chipp}{\chi''}
\begin{document}

\begin{flushright}
  DESY--02--051\\
  DFF 386/04/02\\
  LPTHE--02--22\\
  hep-ph/0204287 \\
  April 2002
\end{flushright}
\begin{center}
\vspace{2.5cm}
{\Large \textbf{\textsf{Tunneling transition to the Pomeron regime
\footnote{Work supported in part by the E.U. QCDNET
 contract FMRX-CT98-0194, MURST (Italy),  by the Polish  
Committee for Scientific Research (KBN) grants no. 2P03B 05119, 2P03B
12019, 5P03B 14420, and by the Alexander von Humboldt Stiftung.}}}}
\vspace{0.7cm}

{\large \textsf{M.~Ciafaloni$^{(a,b)}$,  
D.~Colferai$^{(c)}$,
G.~P.~Salam$^{(d)}$  
and A.~M.~Sta\'sto$^{(a,e)}$}} \\  
\vspace*{0.7cm}
$^{(a)}$ {INFN Sezione di Firenze,  50019 Sesto
  Fiorentino (FI), Italy} \\
\vskip 2mm 
$^{(b)}$ {Dipartimento di Fisica, Universit\'a di Firenze,  50019 Sesto
  Fiorentino (FI), Italy} \\
\vskip 2mm   
{$^{(c)}$II.\ Institut f\"ur Theoretische Physik,  
    Universit\"at Hamburg, Germany}\\  
\vskip 2mm  
$^{(d)}$ {LPTHE, Universit\'es Paris VI et Paris VII, Paris 75005, France} \\  
\vskip 2mm  
$^{(e)}${H. Niewodnicza\'nski Institute of Nuclear Physics,  
 Krak\'ow, Poland} \\   
\vskip 2cm  
\end{center}
\thispagestyle{empty}
\begin{center} {\large \textsf{\textbf{Abstract}}} \end{center}
\begin{quote}
  
  We point out that, in some models of small-$x$ hard processes, the
  transition to the Pomeron regime occurs through a sudden tunneling
  effect, rather than a slow diffusion process. We explain the basis
  for such a feature and we illustrate it for the BFKL equation with
  running coupling by gluon rapidity versus scale correlation plots.

\end{quote}

\begin{center}
PACS 12.38.Cy
\end{center}
\vspace*{2.5 cm}
\newpage

\section{Introduction}

In most situations in QCD, the question of whether a given observable
can be calculated perturbatively reduces to two issues: firstly
whether all the scales in the problem are large compared to
$\Lambda_\mathrm{QCD}$ and secondly whether the observable is infrared and
collinear safe.

One notable exception is the limit of high-energy scattering of two
hard probes, for example quarkonia, or virtual photons. Fixed order
analysis suggests that the scale of $\as$ is simply determined by the
inverse transverse size of the objects being scattered, independent of
the centre of mass energy.  But at very high centre-of-mass energies,
$\sqrt{s}$, it becomes necessary to consider diagrams at all orders in
the coupling.  This resummation was initially considered for fixed
coupling \cite{LLBFKL}, but subsequently interest developed in its
extension to the (more physical) running coupling case
\cite{JK,COLKWIE}. One of the conclusions of these studies was that
regardless of the hardness of the probes, there always exists a centre
of mass energy beyond which the problem becomes entirely
non-perturbative, see for example \cite{Kwiecinski:1999hv,LUND}.

Other next-to-leading corrections \cite{NLLBFKL}, which however must
be resummed, as is done for example in \cite{Salam:1998tj,CC2,CCS2} or
in other related \cite{ABF,FRSV} or less related \cite{BFKLP,SCHMIDT}
approaches, do not qualitatively change this picture. The transition
to the non-perturbative regime takes place also in that case, although
the diffusion effects can be different, due to the lower value of the
intercept and the diffusion coefficient.  This problem is now under
detailed investigation \cite{CCSS2}.

\begin{figure}[htbp]
  \begin{center}
    \epsfig{file=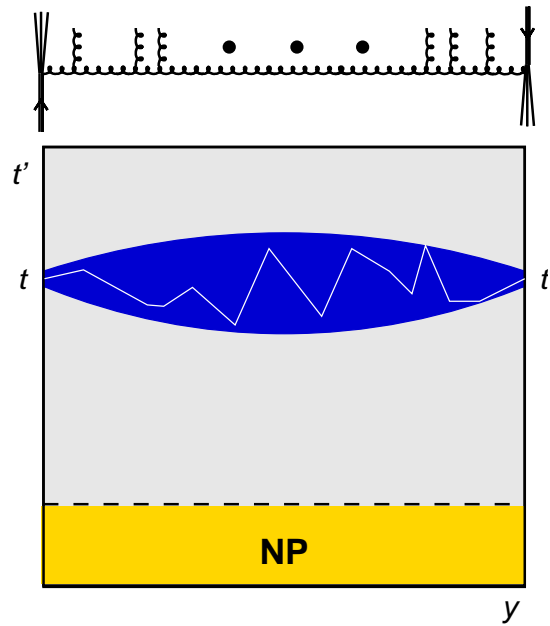}
    \caption{Above: a chain of emissions stretching in rapidity
      ($0<y<Y = \ln s/Q^2$) between the two hard probes at scale
      $t=\ln Q^2/\Lambda^2$. Below a schematic representation of typical
      transverse scales at different stages along the chain: the white
      line illustrates a typical random walk in $t'$ ($\ln k^2$) as
      one goes along the chain, while the dark (blue) `cigar'
      represents the average limits of this diffusion. }
    \label{fig:SchematicDiffusion}
  \end{center}
\end{figure}

For a long time, the view of the transition to the non-perturbative regime was based on the idea of diffusion.  At
lowest order in $\as$ the dominant diagram at high energies is the
exchange of a gluon between the two probes. The scale for $\as$ is
determined by the transverse momentum of the exchanged gluon, $k^2$,
typically of order of transverse scale of the probes, $Q^2$ (in what
follows we will generally refer to $t=\ln Q^2/\Lambda^2$).  At higher
orders one needs to consider emissions of final-state gluons from the
exchanged gluon --- these emissions modify the transverse scale of the
exchanged gluon, which undergoes a random walk in $t'=\ln
k^2/\Lambda^2$. In \cite{Levin:1991xa,Bartels:1993du} this diffusion
was illustrated graphically by plotting the median trajectory in $t'$
as a function of rapidity (the distance along the chain) as well as
the dispersion around that median trajectory. A schematic version of
such a plot is shown in figure~\ref{fig:SchematicDiffusion}, with the
dark (blue) band representing the extent of the diffusion. Its
characteristic shape led to it being named a \emph{cigar}
\cite{Bartels:1993du}.

While at its ends the cigar is restricted to being narrow (by the
presence of the probes), in the middle its width grows as a function
of the centre of mass energy: $\Delta t' \sim \sqrt{\asb Y}$ with $Y=
\ln s/Q^2$. For the problem to remain perturbative, typical
trajectories must remain well outside the non-perturbative infrared
region and this leads to the requirement that the width of the cigar
be less than $t$, \ie $\sqrt{\asb Y } \ll t$. Using $\asb(t) = 1/bt$,
with $b=11/12$ the coefficient of the $1$-loop $\beta$ function, this
reduces to the condition $Y \ll b t^3$. This limit was originally
derived (with some additional numerical factors) in
\cite{Mueller:1997hm,Levin:1995di,Kovchegov:1998ae,Armesto:1998}.

\begin{figure}[htbp]
  \begin{center}
    \epsfig{file=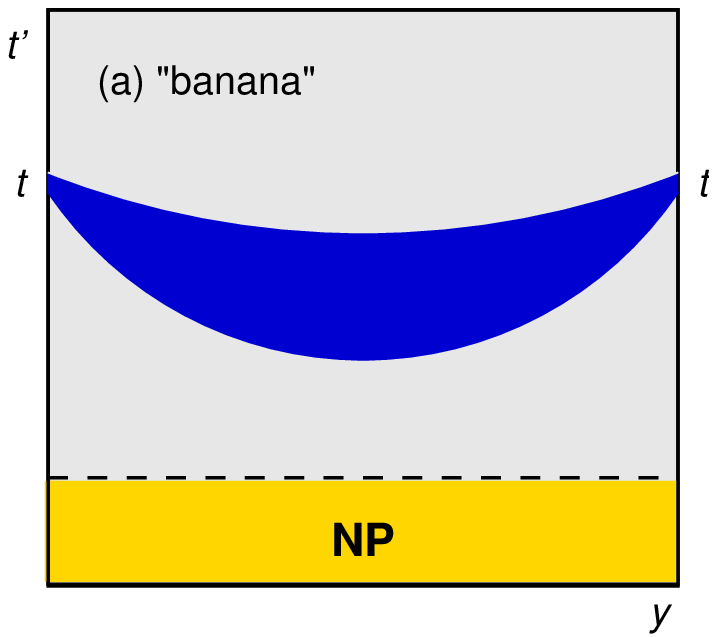, width=0.47\textwidth}
    \hfill
    \epsfig{file=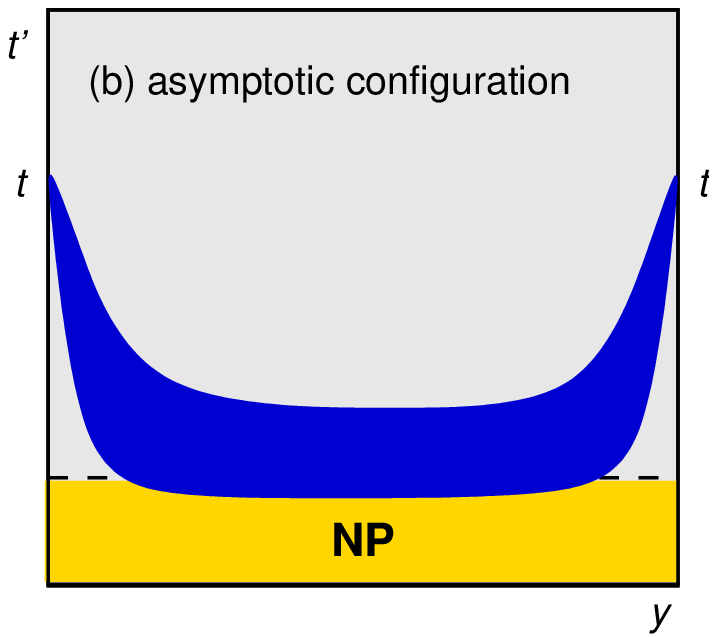, width=0.47\textwidth}
    \caption{Analogues of fig.~\ref{fig:SchematicDiffusion} which show
      the possible effects of the running coupling on the shape of the
      `cigar'.}
    \label{fig:SchematicExtensions}
  \end{center}
\end{figure}

This picture of diffusion has its roots in the fixed-coupling
approximation. It is tempting to consider the full running-coupling
case as a small variation of the fixed-coupling situation. One then
expects the differences to be that the centre of the cigar dips to
$t'$ values below $t$, and that the cigar profile in $t'$ develops
some asymmetry with respect to its centre (the cigar becomes a
`banana', figure~\ref{fig:SchematicExtensions}a). This approach to the
running-coupling case as a perturbation of the fixed-coupling case
leads also to well-defined analytical predictions
\cite{Levin:1995di,Kovchegov:1998ae,Armesto:1998,Ciafaloni:2001db}.
In particular, the usual fixed-coupling exponential growth of the
cross section, $ \log \sigma \sim \as Y$, is supplemented by an extra
term proportional to $b^2 \as^5 Y^3$. This can be interpreted as due
to the centre of the cigar being at $t'$ values of $t -
\order{b \as^3 Y^2}$, which for $Y$ of order $bt^2$ implies that the
cigar reaches the non-perturbative domain. Indeed calculations of the
`purely perturbative' component of the cross section show unphysical
(oscillating) behaviour for $Y \gtrsim b t^2$, which can be
interpreted as meaning that the perturbative component is not a
physically well-defined quantity in that region \cite{Ciafaloni:2001db,CCSS1}.

Recently however, it was discovered in the context of a simplified,
collinear model for BFKL dynamics \cite{Ciafaloni:1999au}, that the
running of the coupling can introduce further qualitative differences.
Rather than the width of the cigar gradually increasing with $Y$ until
it reaches the non-perturbative region at $Y \sim bt^3$, or the centre
of the cigar gradually moving in $t'$ down until it reaches the
non-perturbative region at $Y \sim bt^2$, it turns out that for $Y
\sim t$ an \emph{abrupt} transition occurs and the dominant trajectory
switches from being centred around $t'=t$
(fig.~\ref{fig:SchematicDiffusion}) to being centred around
$t'=\tbar$, where $\tbar \gtrsim 0$ is in the non-perturbative region
(fig.~\ref{fig:SchematicExtensions}b). For such a transition, which we
named `tunneling', there is no passage through any intermediate
`banana' configuration (fig.~\ref{fig:SchematicExtensions}a).

In this article we wish to explain this phenomenon in more detail than
was done previously and for the case of the full BFKL equation with
running coupling.  In section~\ref{sec:theory} we present analytical
calculations which give the estimate of the nonperturbative Pomeron,
while in section~\ref{sec:pictures} we illustrate the phenomenon with
a number of figures resulting from the numerical solution of the BFKL
equation.

\section{Analytical considerations}
\label{sec:theory}

It is useful to recall the form of the BFKL equation as we shall use
it here. We define a BFKL Green's function $G(Y,t,t_0)$ through the
following integro-differential equation
\begin{equation}
\label{eq:Bfkl}
  \frac{dG(Y,t,t_0)}{dY} = \int \frac{d^2 \vec q}{\pi q^2} \,\asb(\ln q^2)
  \left[\frac{k}{k'} G(Y, \ln {k'}^2, t_0) - 
    G(Y, t, t_0) \Theta(k-q) \right]
    , \quad 
  \begin{array}{l}
    k^2 = e^{t}\,,\\
    k'  = |\vec k + \vec q|\,,
  \end{array}
\end{equation}
with the initial condition
\begin{equation}
  \label{eq:deltafn}
  G(0,t,t_0) = \delta(t - t_0)\,,
\end{equation}
and with $\asb = \as N_C / \pi$. All scales are in units of
$\Lambda_\mathrm{QCD}$. We shall take $\asb$ as having its one-loop
perturbative form, regularised by a cutoff at a non-perturbative scale
$\tbar$:
\begin{equation}
\label{eq:asbCutoff}
  \asb(t) = \frac{\Theta(t-\tbar)}{bt}\,,\qquad b = \frac{11}{12}\,.
\end{equation}
In some contexts we will use an equation similar to \eqref{eq:Bfkl},
but with $\as(\ln q^2)$ replaced by $\as(t)$.

It will be convenient to define some high-energy exponents: $\oms$,
the perturbative saddle-point exponent, determines the high-energy
behaviour of the Green's function when it is dominated by scales of
order $t$. It is given by $\oms = (4 \ln2) \asb(t) + \order{\asb^2}$.
We shall also need the non-perturbative pomeron exponent, $\omp$,
which governs the asymptotic high-energy behaviour of the Green's
function, $G(Y) \sim \exp(\omp Y)$, once the problem has become
dominated by non-perturbative scales. It is given by \mbox{$\omp =
  (4\ln2) \asb(\tbar) + \order{\smash{\asb(\tbar)^{5/3}}}$}, though it
is to be kept in mind that the expansion in powers of
$\asb^{5/3}$ converges very slowly.

For the purpose of our arguments we shall take the formal perturbative
limit of $t \gg \tbar$ and correspondingly $ \oms \ll \omp$.

\subsection{Informal arguments for tunneling}

The mechanism of tunneling is illustrated in
figure~\ref{fig:SchematicExtensions}b. To establish when tunneling
takes place we have to equate the contribution from the tunneling
configuration with that from the traditional perturbative
configuration. Neglecting all prefactors, the tunneling configuration
gives a contribution
\begin{equation}
  G_\mathrm{tunnel} \sim e^{-(t-\tbar)/2} \, e^{\omp Y}\,  e^{-(t-\tbar)/2},
\end{equation}
corresponding (reading from left to right) to the `penalty' for
branching from $t$ to $\tbar$, the BFKL evolution at $\tbar$ and the
penalty for branching back to $t$. In contrast, the perturbative
contribution is proportional to
\begin{equation}
  G_\mathrm{pert} \sim  e^{\oms Y}.
\end{equation}
Equating these two expressions gives
\begin{equation}
  \label{eq:YTunnelSimple}
  Y_\mathrm{tunnel}(t) = \frac{t - \tbar}{\omp - \oms(t)}\,.
\end{equation}
Perhaps the most important characteristic of this equation is the fact
that $Y_\mathrm{tunnel}$ is proportional to $t$ rather than $t^3$ as
for the diffusion mechanism.
  One possible exception is the case in which the Pomeron is
weak, that is $\omp \ll 1$. Then $\tbar$ should be quite large, and
including in Eq.(\ref{eq:YTunnelSimple}) the value $\oms(t) =
\chi_m/(bt)$ and the rough estimate $\omp=\chi_m/(b\tbar)$ yields
$Y_\mathrm{tunnel} = (bt/\chi_m)\tbar$, so that the perturbative
regime is kept up to $\oms(t) Y$ values of order $\tbar$.


\subsection{Formal arguments}

For an isolated Pomeron ``bound state'', the tunneling probability
is estimated on the basis of the approximate decomposition
\begin{equation}
G(Y;t,t_0) \; = \; G_{\rm pert}(Y;t,t_0) + e^{\omp Y} \frac{\CF_{\omp}(t) \CF_{\omp}(t_0)}{\int dt {\CF_{\omp}(t)}^2} \, ,
\label{eq:NormSymm}
\end{equation}
where $\CF_{\omp}(t)$ is the regular solution for $t\rightarrow
\infty$ of the BFKL equation \cite{CC1}, satisfying the
strong-coupling boundary condition also.  The decomposition
(\ref{eq:NormSymm}) separates the continuum from the lowest bound
state contributions to the gluon's Green functions \cite{CCSS1} or,
equivalently, the background integral from the leading Regge pole
\cite{Ciafaloni:2001db}. The normalisation of the bound state factor
in front of $e^{\omp Y}$ is simply obtained by imposing that it be a
projector by $t_0$ integration, or equivalently from
the requirement that
\begin{equation}
  G(Y;t,t_0) = \int dt' G(Y;t,t') G(Y-y;t',t_0)\,.
\end{equation}
Eq.(\ref{eq:NormSymm}) applies as it stands to a
symmetrical BFKL kernel, as that with the scale choice $\asb(\ln q^2)$
in Eq.(\ref{eq:Bfkl}). If instead the scale $\asb(t)$ is chosen, the
kernel is asymmetrical, the conjugated wave function is
$\bar{\CF}_{\omp}(t)=t \CF_{\omp}^*(t)$, and the bound state
projectors differ by a $t_0$ factor and by the normalisation integral
\begin{equation}
G(Y;t,t_0) \; = \; G_{\rm pert}(Y;t,t_0) + e^{\omp Y} \frac{\CF_{\omp}(t) \CF_{\omp}(t_0) t_0 }{\int dt \, t \, {\CF_{\omp}(t)}^2} \, .
\label{eq:NormASymm}
\end{equation}
In either case, the large $t$ behaviour of $\CF_{\omp}(t)$ is
estimated in a semiclassical approach in which $b\asb(t), 1/Y$ and
also $b\omp$ are taken to be small parameters (weak Pomeron, or $b$ -
expansion \cite{CCSS1}).  In such a case the $\gamma$ -
representation of $\CF_{\omp}$ is valid, and takes the form
\begin{multline}
\CF_{\omp}(t)  =  \int_{1/2-i \infty}^{1/2+i \infty}
\frac{d\gamma}{2\pi i} e^{(\gamma-1/2)t - \frac{1}{b\omp}
  X_{\omp}(\gamma)} \\ \simeq
\sqrt{\frac{b\omp}{-2\pi \chi'[p(b\omp t)]}}\exp\left[-\frac{1}{b\omp}
  \int_{\chi_m}^{b\omp t} p(x) \, dx \right] \; ,
\label{eq:SemiClas}
\end{multline}
where the saddle point condition  at $\gambar = \frac{1}{2} - p $
\begin{equation}
\chi\left(\frac{1}{2}-p\right) \; = \; \frac{\partial
  X(\gamma)}{\partial \gamma}
\bigg|_{\gamma=\frac{1}{2}-p} \; = \; b\, \omp t \; ,
\label{eq:SaddlePoint}
\end{equation}
defines the semiclassical ``momentum'' $p(b\omp t)$.  The effective
eigenvalue function in Eq.(\ref{eq:SaddlePoint}) is just the leading
characteristic function $\chi_0(\gamma)$ for the $\asb(t)$ choice, and
differs by some corrections involving the parameter $\omp$ in the
symmetric scale choice $\asb(\ln q^2)$ which essentially lead to a
rescaling of the (right) regular solution by a constant factor $\sim
1/\sqrt{\chi_m}$ (see Appendix) .

The Pomeron suppression arising from Eq.(\ref{eq:SemiClas}) has a
universal form $\sim e^{\left(-\frac{1}{b\asb(t)} g(\asb(t))
\right)}$, where the function $g(\asb)$ ($g(0) \neq 0$) is easily found
from Eq.(\ref{eq:SaddlePoint}), \cite{CCSS1}.  On the other hand the
perturbative part, corresponding to the continuum, is dominated by the
saddle point exponent $\oms(t)$
\begin{equation}
G_{\mathrm{pert}}(t,t_0;Y) \; = \; \frac{1}{\sqrt{2\pi \chi_m^{''}
    \asb(\frac{t+t_0}{2}) Y}} \exp\left(\oms\left(\frac{t+t_0}{2}\right) Y +
\mathrm{corrections}\right) \; ,
\label{eq:GreenPert}
\end{equation}
where the diffusion corrections are due to the running coupling
\cite{Levin:1995di,Kovchegov:1998ae,Armesto:1998}.

From Eqs.(\ref{eq:NormSymm},\ref{eq:NormASymm},\ref{eq:SemiClas}) we estimate the $t$-dependence of the Pomeron contributions for the choices $\asb(k^2)$ ($\asb(q^2)$) of the running coupling scale respectively
\begin{eqnarray}
\label{eq:GPomEstimate}
G_{\pom}^{(k)} & = & \frac{N^{(k)}(\tbar)}{2\pi b \omp t} 
\exp\left[-t+\frac{2}{\omp}\left(\log b\omp t + 1 + \psi(1) + \order{\frac{1}{(b\omp t)^3}}\right) \right] \, , \nonumber \\
G_{\pom}^{(q)} & = & \frac{N^{(q)}(\tbar)}{2\pi (b \omp t)^2}
\left(\frac{b\omp}{\chi_m} \right) 
\exp\left[-t+\frac{2}{\omp}\left(\log b\omp t + 1 + \psi(1) + \order{\frac{1}{(b\omp t)^3}}\right) \right] \, . \nonumber \\
\end{eqnarray}
Notice that they differ both in the $t$-dependence, because of the $t$ factor on Eq.(\ref{eq:NormASymm}) and for the $\tbar$-dependent normalisation integrals
\begin{eqnarray}
N^{(k)}(\tbar) = \left(\int_{\tbar}^{\infty} dt \,  t \, \CF_{\omp}^2(t)
 \right)^{-1} \, ,\nonumber \\
N^{(q)}(\tbar) = \left(\int dt  \CF_{\omp}^2(t) \right)^{-1} \, ,
\label{eq:NormIntegral}
\end{eqnarray}
The latter quantities cannot be estimated in the semiclassical approximation, which breaks down at the turning points, and they need to be evaluated numerically. We can roughly estimate their size on the basis of the Airy diffusion model
\cite{CC1} obtained by the quadratic expansion of the characteristic
function
\begin{equation}
\chi_{\om}(\gamma) \;= \; \chi_m \; + \; \frac{1}{2} \chi_m''
(\gamma-\gamma_m)^2 + \cdots \; .
\label{eq:airy}
\end{equation}
This approximation is expected to be a reliable one for the Pomeron
dynamics provided the fluctuations $\langle(\gamma-\gamma_m)^2\rangle \sim
\frac{2b\omp}{\chi''_m}$ are small compared to $4!\chi''_m/(2
\chi_m^{(4)})$, and $t$ varies in a range for which
\begin{equation}
\xi_{\omp}(t) \;= \; \left( \frac{2b
    \omp}{\chi_m''}\right)^{\frac{1}{3}}\left(t-\frac{\chi_m}{b\omp}
\right) \; ,
\end{equation}
is a parameter of order unity.  In this case the BFKL equation
with the choice $\asb(k^2)$ 
reduces, by Eq.(\ref{eq:airy}) to a second-order differential
equation in $t$ \cite{CC1}, and the cut-off condition for $\asb(t)$ at
$t=\tbar$ (\ref{eq:asbCutoff}) implies that the regular solution for
$t\rightarrow -\infty$ must vanish at $\tbar$ , and for $t$ around
$\tbar$ takes the following form
\begin{equation}
\label{eq:SolAiry}
\left(\frac{\chipp_m}{2b\omega}
\right)^{\frac{1}{3}}\tilde{\CF}_{\om}(t) \; = \; \pi \left[
\mathrm{Bi}[\xi_{\om}(t)] -
 \frac{\mathrm{Bi}
  [\bar{\xi}_{\om}(\tbar)]}{\mathrm{Ai}[\bar{\xi}_{\om}(\tbar)]}
\mathrm{Ai}[\xi_{\om}(t)] \right] \; ,
\end{equation}
while the regular solution for $t \rightarrow \infty$ becomes
\begin{equation}
\left(\frac{\chi''_m}{2 b \omega}\right)^{\frac{1}{3}}
\CF_{\omega}(t) = \mathrm{Ai}\left[\xi_{\omega}(t) \right] \; ,
\label{eq:RegSolAiry}
\end{equation}
with relative normalisation fixed by the Wronskian
\begin{equation}
\mathrm{W}(\CF_{\omega},\tilde{\CF}_{\omega}) = \frac{2 b \omega}{\chi''_m} \; .
\label{eq:Wronskian}
\end{equation}

Using now the Airy form for the regular solution (\ref{eq:RegSolAiry}) we can evaluate the normalisation integrals (\ref{eq:NormIntegral}) to get
\begin{eqnarray}
\label{eq:NIntegralAiry}
N^{(k)}(\tbar) &  =  & \frac{\chi_m}{b\omp} \left( \frac{2 b\omp}{\chi_m^{''}}\right)^{\frac{1}{3}} \int_{-\xi_0}^{\infty} \!\! d\xi [\mathrm{Ai}(\xi)]^2+\int_{-\xi_0}^{\infty} \!\! d\xi \xi [\mathrm{Ai}(\xi)]^2 = \nonumber \\
& =  &  
\xi_0 [\mathrm{Ai}^{'}(-\xi_0)]^2 \frac{\frac{2}{3}\oms(\tbar) + \frac{1}{3}\omp}{\oms(\tbar)-\omp}\,, \nonumber \\
\chi_m N^{(q)}(\tbar) &  = & \left(\frac{2 b \omp}{\chi_m^{''}}\right)^{\frac{1}{3}} 
 \int_{-\xi_0}^{\infty} \!\! d\xi [\mathrm{Ai}(\xi)]^2 = \xi_0 \frac{\oms(\tbar)}{\oms(\tbar)-\omp} \frac{b\omp}{\chi_m}[\mathrm{Ai}^{'}(-\xi_0)]^2\,,
\end{eqnarray}
where we have used the Airy identities \cite{GRADRYZHIK} and the assumption
that the Pomeron intercept is given by the rightmost zero of the Airy $\mathrm{Ai}(\xi)$ function
\begin{equation}
\left(\frac{2b\omp}{\chi_m''} \right)^{\frac{1}{3}} \left(\tbar -
  \frac{\chi_m}{b\omp} \right) \; =\; -\xi_0 \simeq -2.3381 \; ,
\hspace{1cm} \mathrm{Ai}(-\xi_0) =0 \; .
\label{eq:PomInter}
\end{equation}
The additional factor $\chi_m$ which multiplies $N^{(q)}(\tbar)$ comes from 
the rescaling of the (right)regular eigenfunctions due to a scale change
$\CF^{(q)}_{\omp}(t) \sqrt{\chi_m} \simeq \CF^{(k)}_{\omp}(t)$.

The expressions (\ref{eq:GPomEstimate}) supplemented by the normalisation
factors (\ref{eq:NIntegralAiry}) can be compared with the naive estimate
(\ref{eq:YTunnelSimple}) by rewriting them in the form
\begin{eqnarray}
\label{eq:GreenPom2}
G_{\mathrm{I\!P}}^{(k)} & = & \asb(t) \exp\left[-(t-\tbar) +
  \frac{2}{b\omp}\log\frac{t}{\tbar} + \Delta^{(k)} \right] \; , \nonumber \\
G_{\mathrm{I\!P}}^{(q)} & = & \asb^2(t) \exp\left[-(t-\tbar) +
  \frac{2}{b\omp}\log\frac{t}{\tbar} + \Delta^{(q)} \right] \; ,
\end{eqnarray}
where one sees the explicit $\asb(t)$, ($\asb(t)^2$) dependence 
for $\asb(k^2)$ and $\asb(q^2)$ scale choices respectively.
The $\log \frac{t}{\tbar}$ in the exponent represents a collinear
enhancement due to the evolution from scale $t$ to scale $\tbar$, and
the $\tbar$ - dependent factors 
\begin{eqnarray}
\label{eq:Delta}
\exp(\Delta^{(k)}) & = & \frac{1}{2 \pi \xi_0 [\mathrm{Ai}'(-\xi_0)]^2} \;
\frac{ [\oms(\tbar)-\omp]}{\frac{1}{3}\omp + \frac{2}{3} \oms(\tbar)}
\; \frac{1}{\omp} \exp\left[\frac{2}{b\omp}(1+\psi(1)+\log b\omp\tbar)
  - \tbar \right] \; , \nonumber \\
\exp(\Delta^{(q)}) & = & \frac{\chi_m}{2 \pi \xi_0 [\mathrm{Ai}'(-\xi_0)]^2} \;
\frac{ [\oms(\tbar)-\omp]}{\oms(\tbar) \, \omp^2}
\;  \exp\left[\frac{2}{b\omp}(1+\psi(1)+\log b\omp\tbar)
  - \tbar \right] \; , 
\end{eqnarray}
come from the normalisation factors (\ref{eq:NIntegralAiry}) and of the
semiclassical prefactors. By then comparing the result
(\ref{eq:GreenPom2}) with the perturbative Green's function in
Eq.(\ref{eq:GreenPert}) we obtain the estimate of the tunneling
transition rapidity in both scale choice cases
\begin{align}
(\omp-\oms(t)) \, Y_\mathrm{tunnel}^{(k)}  & = t-\tbar - \frac{2}{b\omp}
\log\frac{t}{\tbar} - \log \asb(t) -\frac{1}{2} \log(2\pi \chipp_m
\asb(t)Y_\mathrm{tunnel} ) - \Delta^{(k)}(\tbar) \; , \nonumber \\
(\omp-\oms(t)) \, Y_\mathrm{tunnel}^{(q)}  & = t-\tbar - \frac{2}{b\omp}
\log\frac{t}{\tbar} - 2 \log \asb(t) -\frac{1}{2} \log(2\pi \chipp_m
\asb(t)Y_\mathrm{tunnel} ) - \Delta^{(q)}(\tbar) \; , \nonumber \\
\label{eq:YTunnelExact}
\end{align}

\subsection{Limits of the tunneling picture}

We have just seen that the Pomeron regime sets in, according to
Eqs.(\ref{eq:YTunnelSimple}) and (\ref{eq:YTunnelExact}) at $\oms(t)
Y$ values of order $\tbar$. This is to be compared with the validity
boundary of the hard Pomeron behaviour \cite{CCSS1} $1 \lesssim
\oms(t) Y \lesssim t / \sqrt{D}$, where $D = \frac{\chipp_m}{2\chi_m}$
is the diffusion coefficient. Thus, for phenomenological values of $t$
and $\tbar$ there appears to be little room for the hard Pomeron (and
its diffusion corrections) to be observable, at least for models (such
as the leading BFKL equation) in which the hard Pomeron and the
nonperturbative Pomeron are quite strong. The Pomeron contamination in
such cases, is a serious problem even for a theoretical determination
of perturbative diffusion corrections to the hard Pomeron. In fact,
one has to use the $b$-expansion \cite{CCSS1} --- $b \rightarrow 0 $
limit with $\asb(t)$ and $\asb(\tbar)$ fixed --- in order to identify
such corrections at the numerical level.

The situation may change, if subleading and unitarity corrections are
taken into account. Subleading contributions to the kernel are large
\cite{NLLBFKL} and need to be resummed \cite{Salam:1998tj,CC2,CCS2}:
they provide eventually a sizeable decrease of the hard Pomeron
intercept and even more of the diffusion coefficient.  This effect
slows down both the nonperturbative effects and diffusion
corrections, thus increasing 
the range of validity of the perturbative predictions. Furthermore,
unitarity effects are expected to be important for large densities
and more so in the strong coupling region. All this goes in the
direction of a weaker asymptotic Pomeron --- as is the actual physical
Pomeron of soft physics \cite{DL} --- thus increasing the
effective $\tbar$ value, and the window of validity of the hard
Pomeron regime.\footnote{Indeed if the soft Pomeron has weaker energy
  dependence than the hard pomeron, then tunneling stops being a
  favoured contribution altogether since the non-perturbative Pomeron
  cannot overtake the hard pomeron before the oscillatory regime is reached.}

It may thus be that in realistic small-$x$ models \cite{CCSS2} with
a weaker Pomeron, diffusion corrections to the hard Pomeron may set in
before tunneling, so that the transition to the Pomeron regime is
smoother. In any case, before diffusion corrections lead to a
decreasing cross-section and to the asymptotic oscillatory
regime\cite{Ciafaloni:2001db} the Pomeron is expected to take over.

\section{Pictorial results}
\label{sec:pictures}

\subsection{Cigars}

Perhaps the most dramatic illustration of the phenomenon of tunneling
can be obtained through pictorial representations analogous to those
of figures~\ref{fig:SchematicDiffusion} and
\ref{fig:SchematicExtensions}, but obtained by exact numerical solution 
of the BFKL equation.

This involves studying the running-coupling dynamics of BFKL evolution
and in particular the sensitivity of the Green's function $G(Y,t,t)$
to evolution at some intermediate point $y,t'$. This can be done by
examining:
\begin{equation}
 f(Y,y,t,t') = \frac{G(y,t,t') G(Y-y,t',t)}{G(Y,t,t)}\,,
\end{equation}
which, by virtue of the linearity of the BFKL equation, is normalised
as follows
\begin{equation}
  \int dt' f(Y,y,t,t') = 1\,.
\end{equation}
From the point of view of graphical representation this quantity
suffers from the problem that for $y\simeq0$ or $y\simeq Y$ and
$t'\simeq t$ one is sensitive to the $\delta$-function initial
condition for $G$, eq.~\eqref{eq:deltafn}. To smooth-out this
$\delta$-function we examine instead a smeared version of $f$:
\begin{equation}
  \label{eq:smear}
 \fbar (Y,y,t,t') = 
\frac{\int dt_0\, dt_1 \, e^{-(t_0-t)^2/2-(t_1-t)^2/2}
  \, G(y,t_0,t') G(Y-y,t',t)}{
  \int dt_0\, dt_1 \, e^{-(t_0-t)^2/2-(t_1-t)^2/2} \, G(Y,t_0,t_1)}
\end{equation}

\begin{figure}[htbp]
  \begin{center}
    \epsfig{file=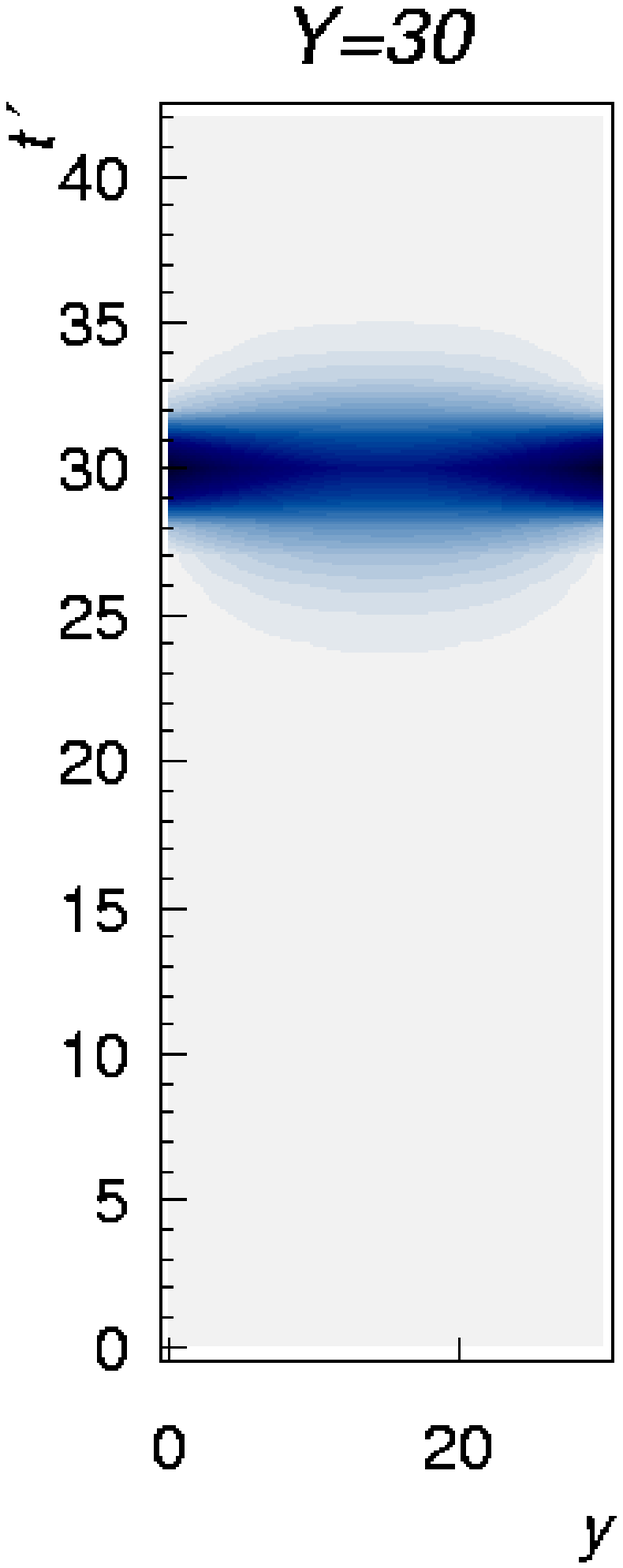,height=0.25\textheight}
    \epsfig{file=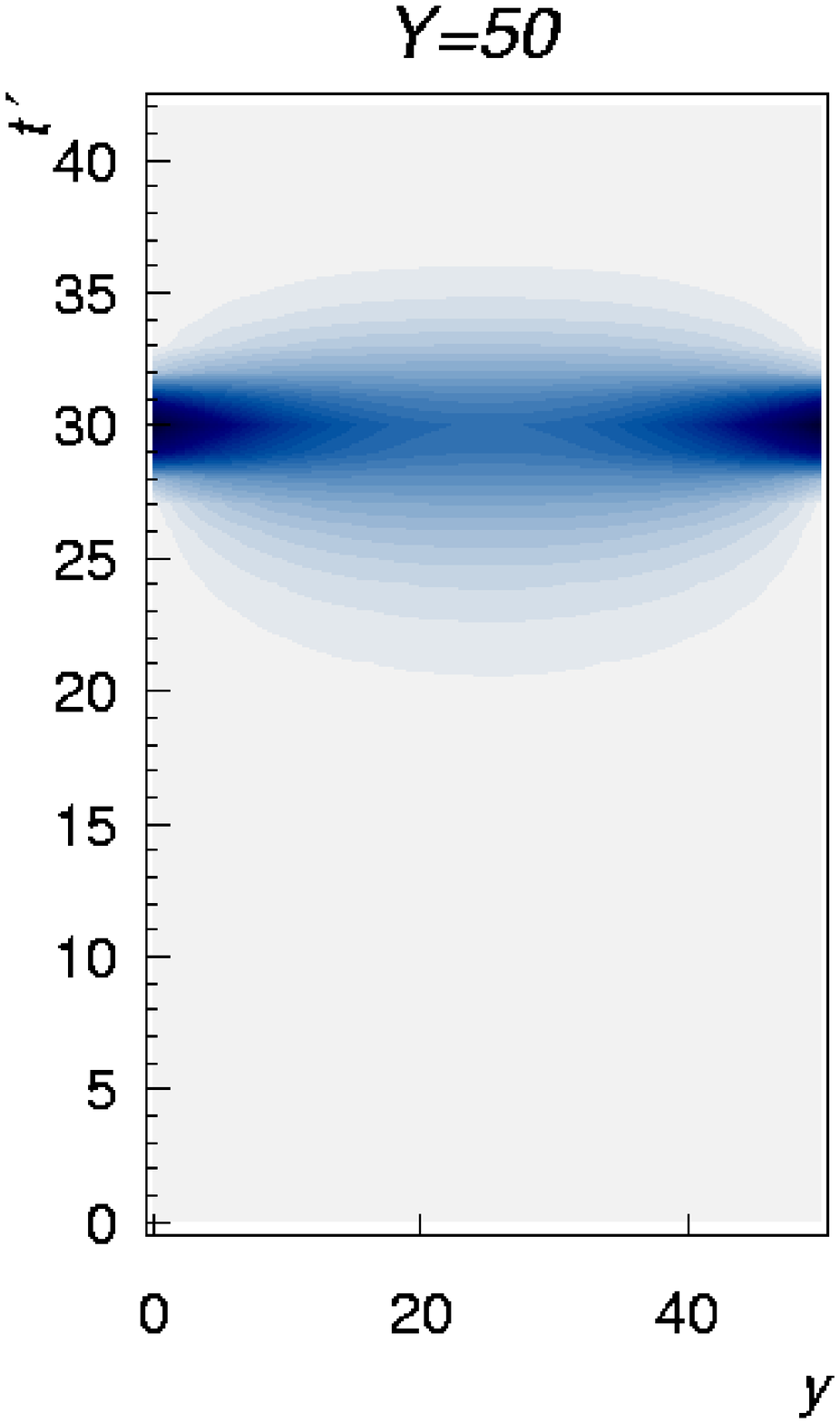,height=0.25\textheight}
    \epsfig{file=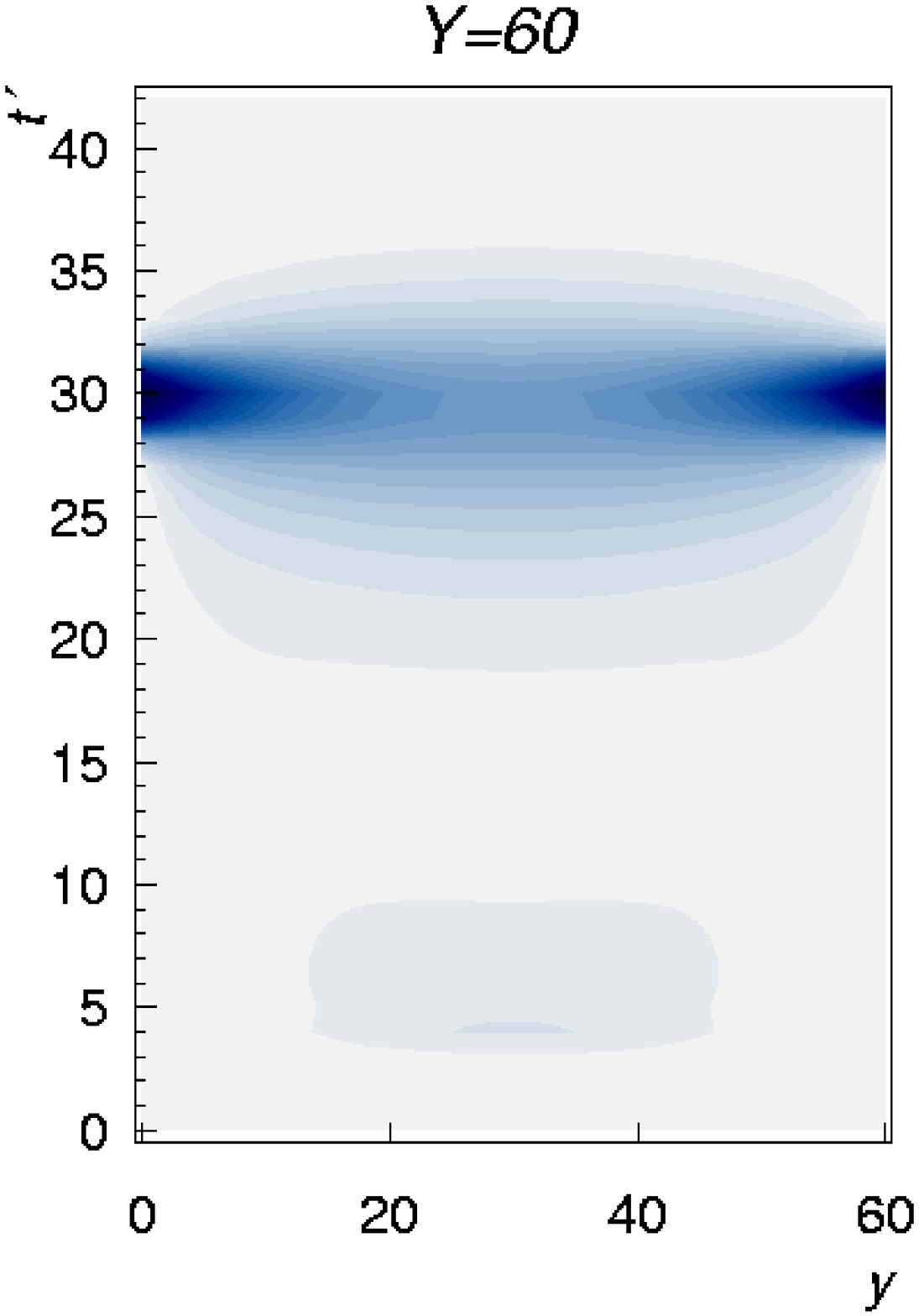,height=0.25\textheight}
    \epsfig{file=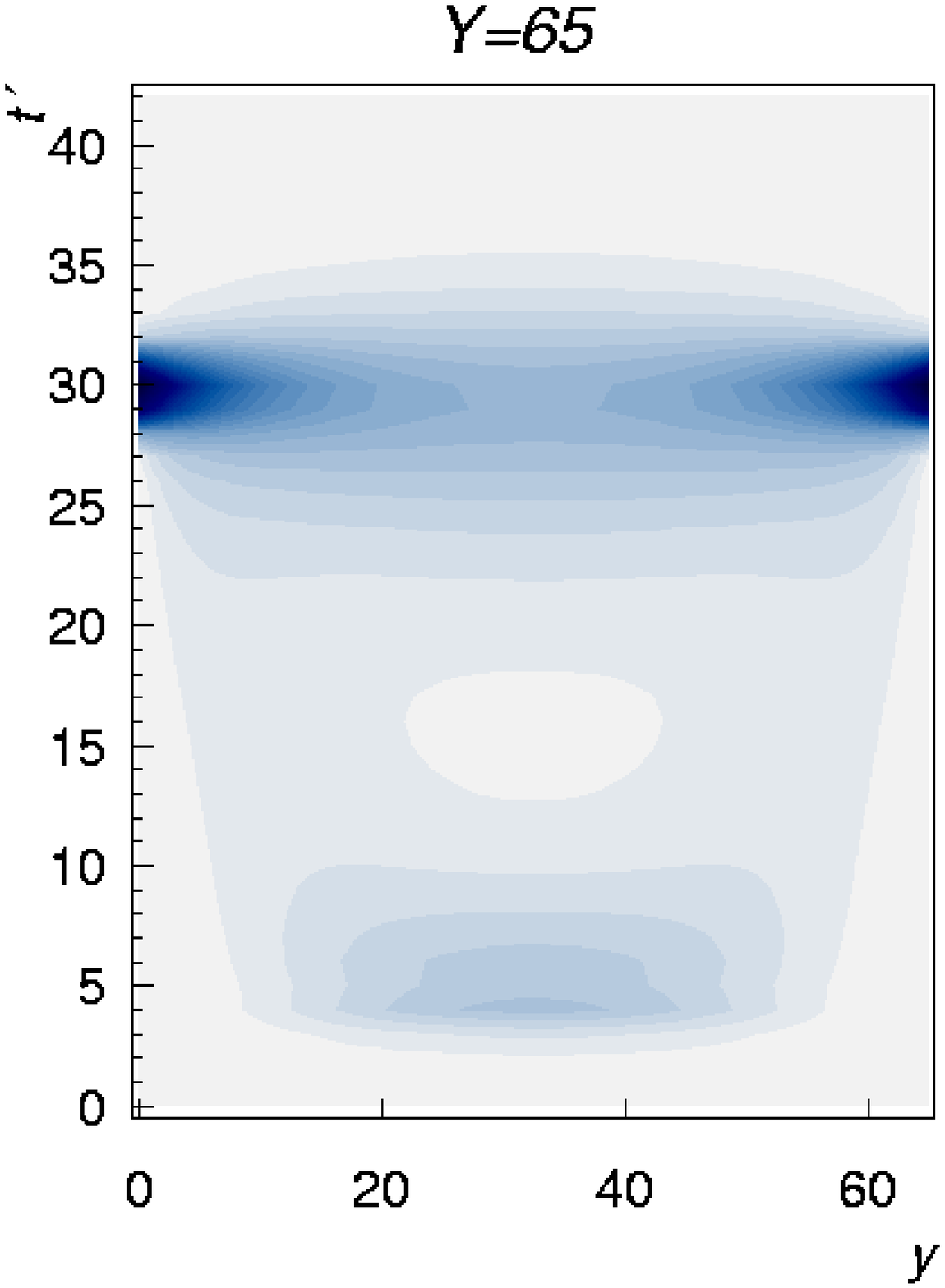,height=0.25\textheight}
    \vspace{0.2cm}\\
    \epsfig{file=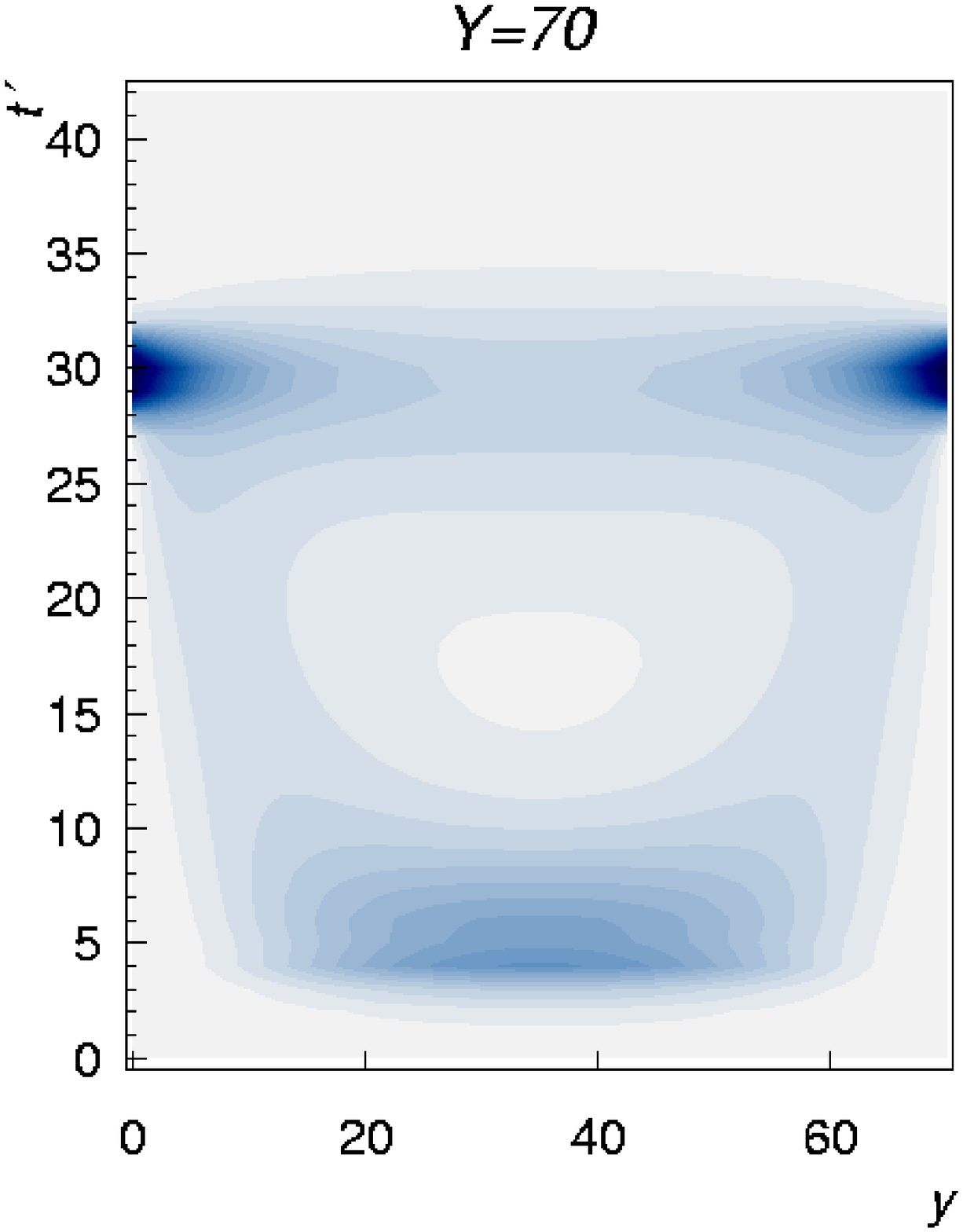,height=0.25\textheight}
    \epsfig{file=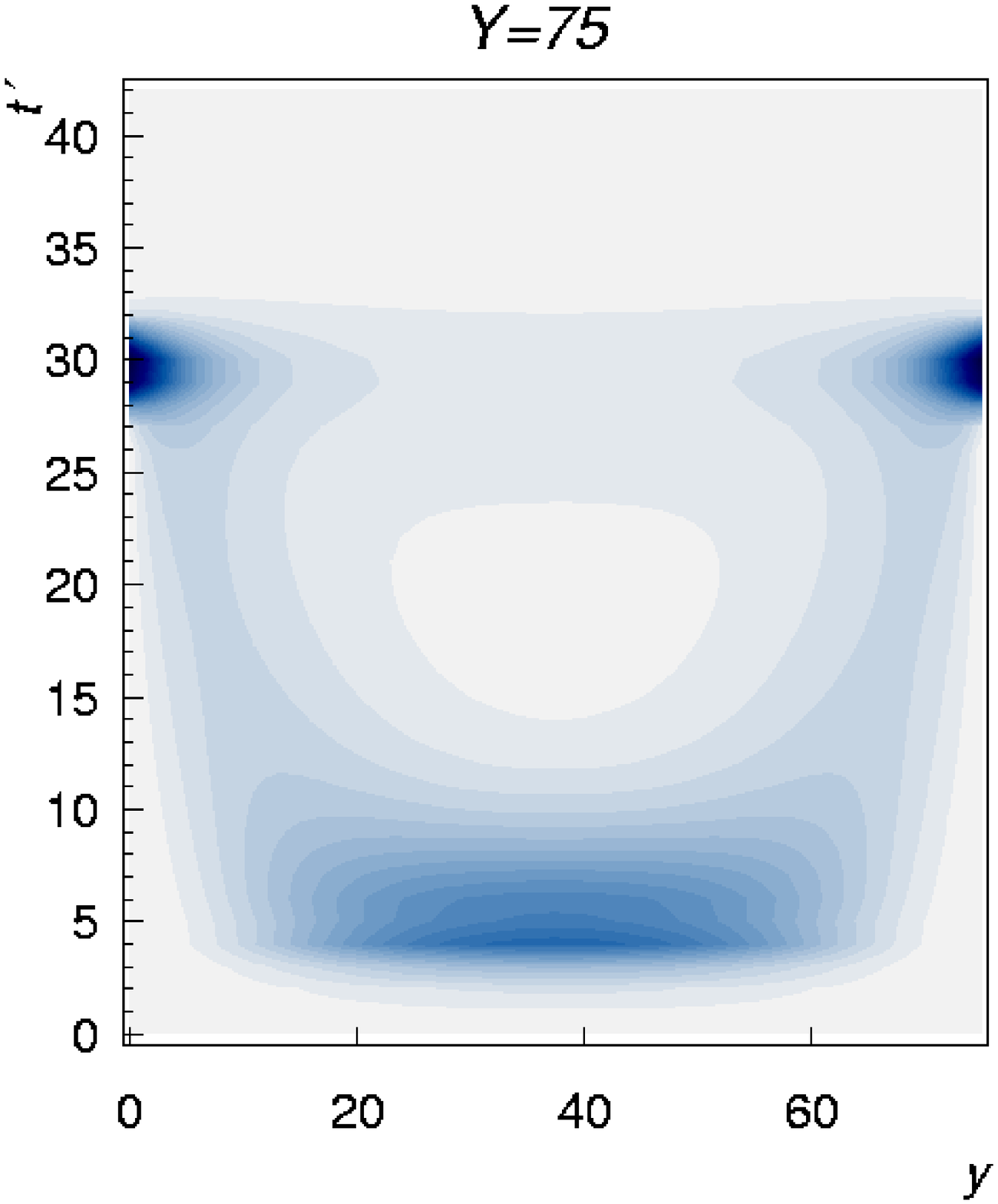,height=0.25\textheight}
    \epsfig{file=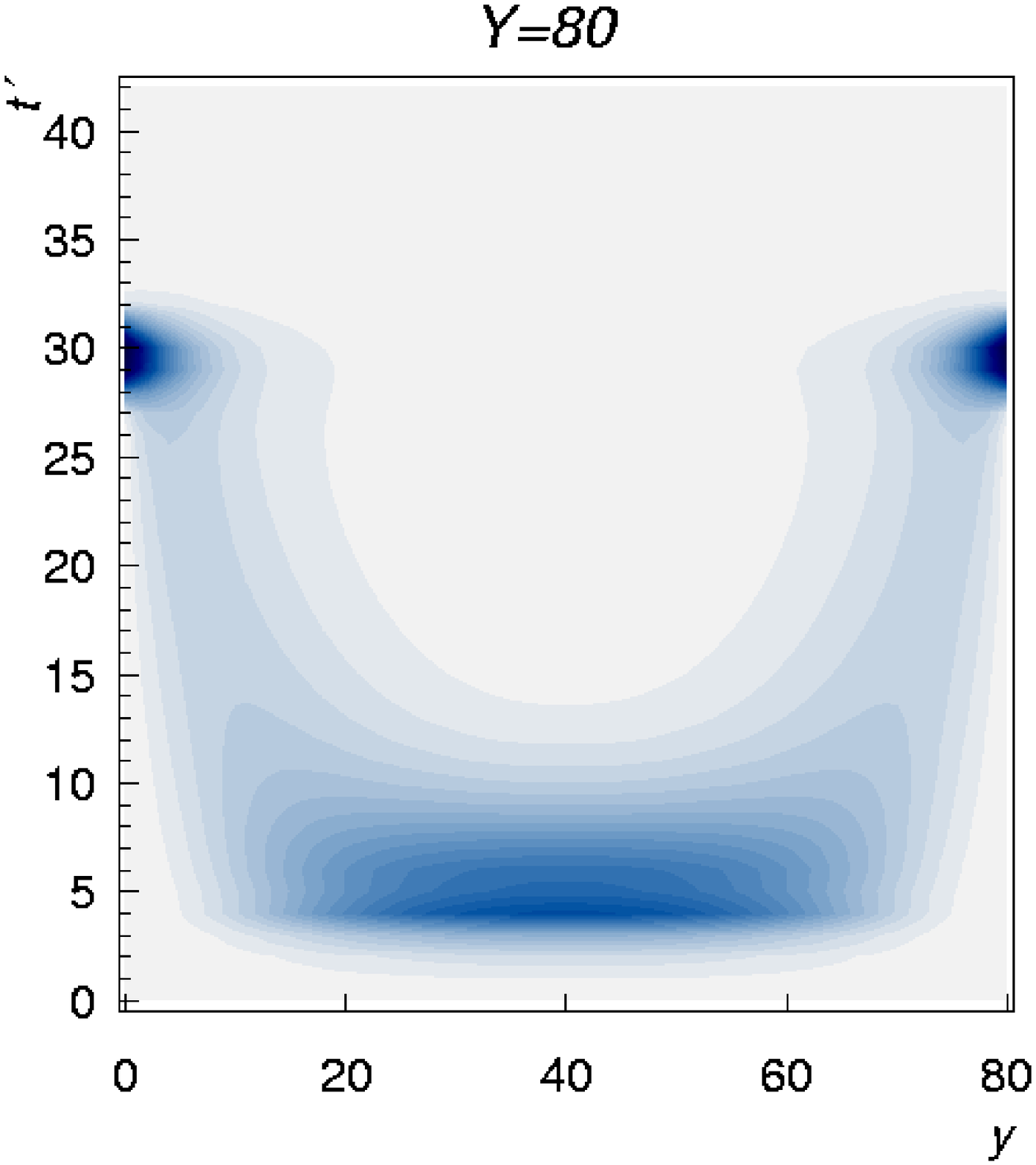,height=0.25\textheight}
    \vspace{0.2cm}\\
    \epsfig{file=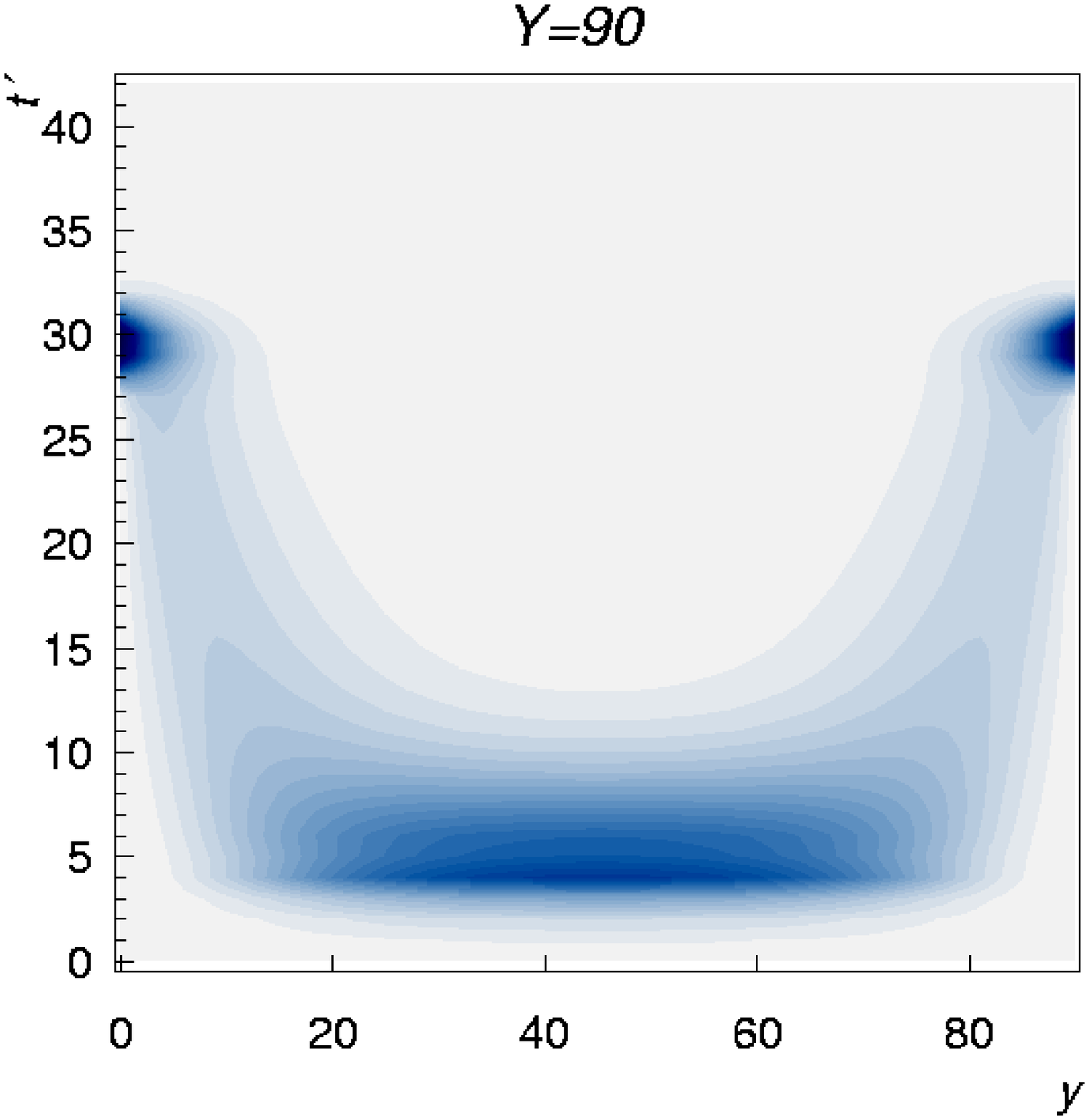,height=0.25\textheight}
    \epsfig{file=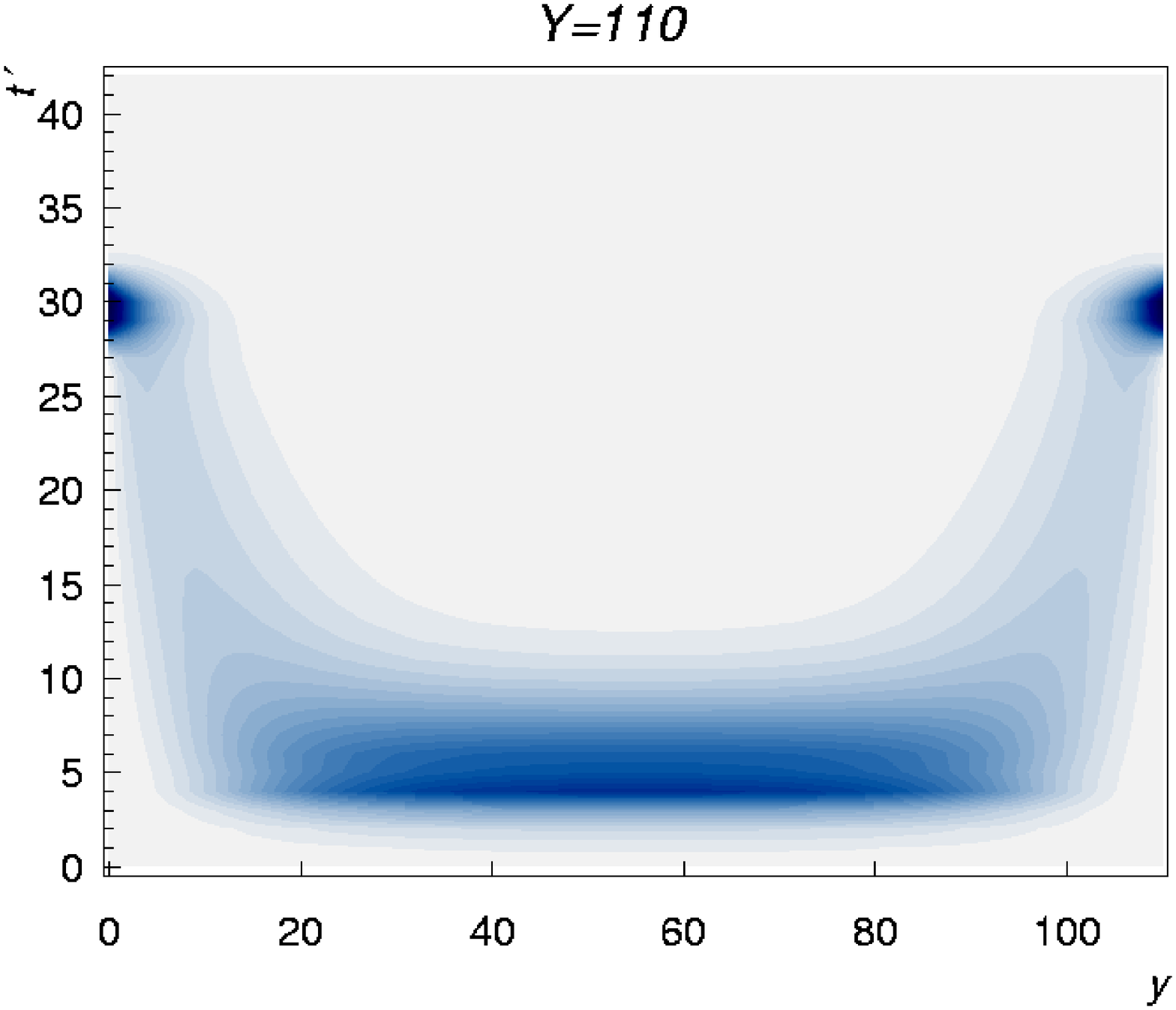,height=0.25\textheight}
    \caption{Contour plots for $\fbar(Y,y,t,t')$, illustrating different
      stages of the evolution: $Y=30$ and $50$ illustrates standard
      `cigar' type plots, $Y=60$--$70$ show the point where tunneling
      begins to play a role (BFKL `aliens'), while for $Y\gtrsim80$
      tunneling has become the dominant contribution (BFKL `bowls').
      MPEG animations of the transition are available from
      \texttt{http://www.lpthe.jussieu.fr/\~~\!\!\!\!salam/tunneling/}
      which includes additionally results for other
      combinations of $t$ and $\tbar$.}
    \label{fig:cigars}
  \end{center}
\end{figure}

Contour plots of $\fbar(Y,y,t,t')$ as a function of $y$ and $t'$ are
shown in figure~\ref{fig:cigars} for various values of $Y$. We have
chosen a value for $t$ that is physically unrealistic, $t=30$.  The
reason for this choice is that it enhances the separation between the
perturbative and non-perturbative regions, allowing us to illustrate
more clearly the transition that takes place.\footnote{For more
  phenomenologically relevant plots we note that in addition to taking
  more reasonable values for $t$ it would also be necessary to include
  the NLL corrections and collinear resummation.}  The coupling is
regularised in the infrared, eq.~\eqref{eq:asbCutoff}, by a cutoff at
$\tbar=4$.

At the lowest value of $Y$ in figure~\ref{fig:cigars} the shape of the
cigar is largely determined by the smearing of the initial condition,
eq.~\eqref{eq:smear}. At $Y=50$ one can see a significant amount of
diffusion at central values of $y$, as well as a slight asymmetry
favouring lower $t'$ values, caused by the running of the coupling.

At $Y=60$ the tunneling transition starts to take place: at low values
of $t'\simeq t$, an enhanced region appears which is
\emph{disconnected} from the main region of the cigar --- it has not
appeared through diffusion, but through tunneling. As $Y$ increases
the perturbative region of $t'\simeq t$ becomes progressively less
important, eventually being completely supplanted by the
non-perturbative region 
$t'\simeq \tbar$. In this transition stage we take the liberty of
referring to the contour plots as `BFKL aliens'.

Beyond $Y=75$, the evolution takes place entirely in the
non-perturbative region, except in the wings, $y\simeq0$ and $y\simeq
Y$, which connect the initial conditions at $t$ to the infrared
region. We refer to these contour plots as `bowls'.

We note that the position in $Y$ of the tunneling transition is quite
consistent with the arguments of section~\ref{sec:theory}.  For our
particular non-perturbative conditions, the non-perturbative power is
$\omp \simeq 0.32$; substituting this into
eq.~\eqref{eq:YTunnelExact} gives $Y_\mathrm{tunnel} \simeq 68$.

\subsection{Ducks}

\begin{figure}[htbp]
  \begin{center}
    \epsfig{file=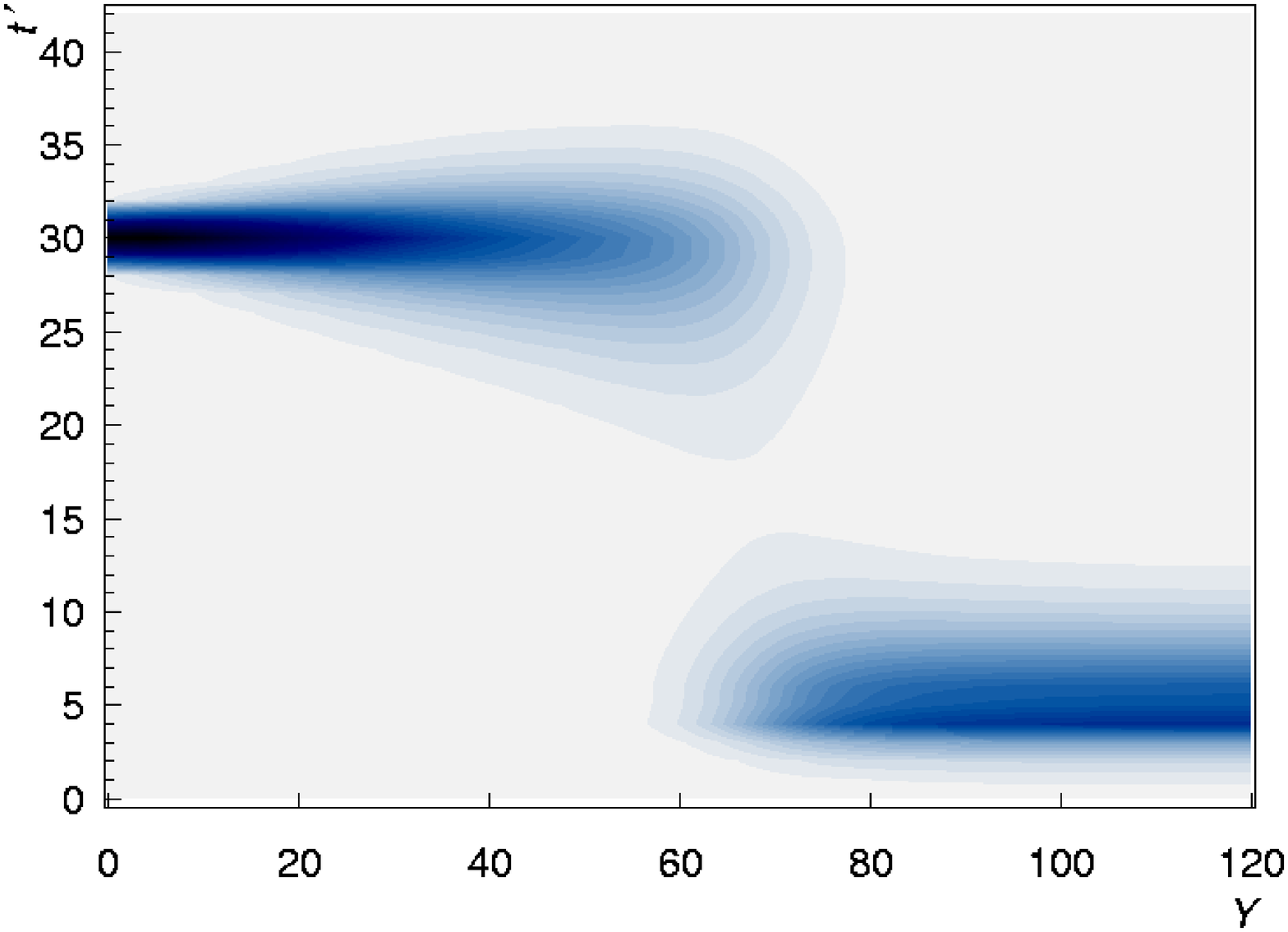,height=0.45\textwidth}\vspace{0.3cm}\\
    \mbox{ }\,\epsfig{file=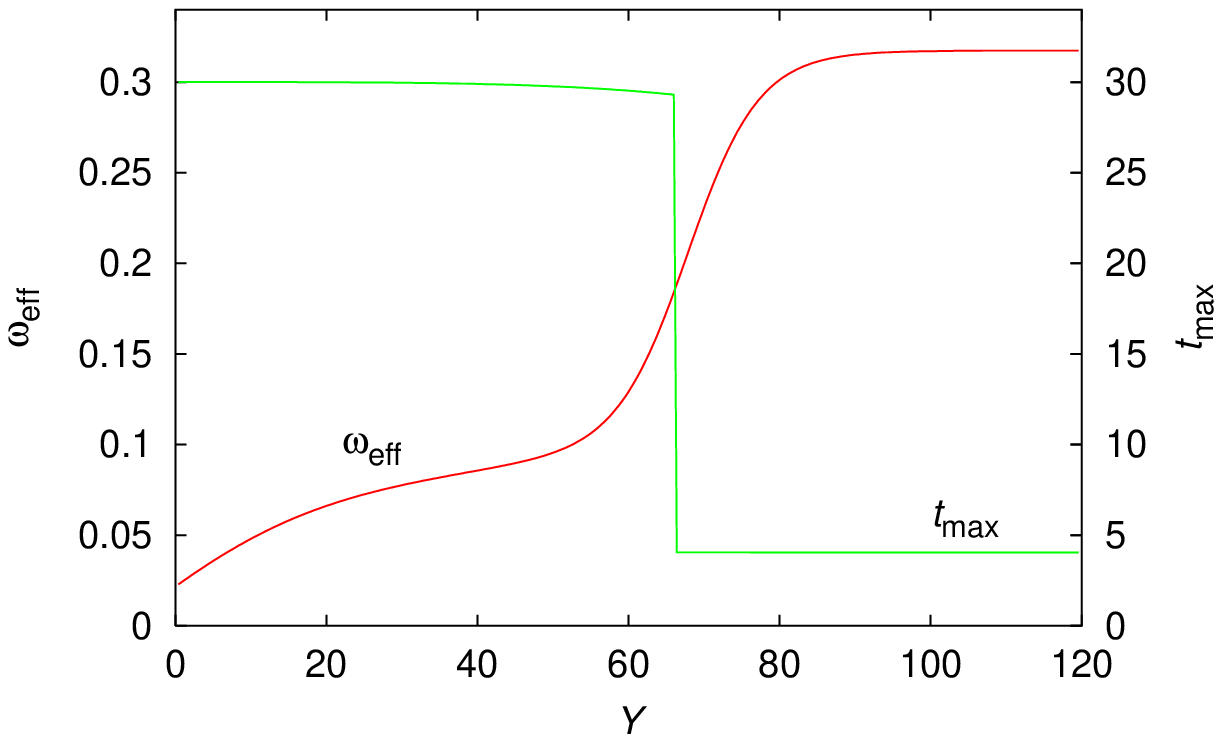,height=0.448\textwidth}
    \caption{The upper figure is a contour plot of
      $\fbar (Y,Y/2,t,t')$, corresponding to a cross section at $Y/2$
      of `cigar'-type plots in fig.~\ref{fig:cigars}, as a function of
      $Y$. We refer to the resulting figure as a `duck'.  The lower
      plot shows the effective exponent, $\omega_\mathrm{eff}$, and
      the position of the maximum of $\fbar$, $t_\mathrm{max}$, as a
      function of $Y$.}
    \label{fig:ducks}
  \end{center}
\end{figure}

A convenient way of visualising the tunneling transition within a single
plot is to consider $\fbar (Y,Y/2,t,t')$ as a function of $Y$ and
$t'$, \ie a contour plot built up 
of cross sections through the centres of the plots of
figure~\ref{fig:cigars}, for a range of $Y$ values. This is shown in
the upper plot of figure~\ref{fig:ducks}. At low $Y$ we see just the
width of the initial condition, which gradually diffuses out while
remaining centred at $t'\simeq t$. At $Y\simeq 60$ a second region
appears `out of nowhere' at $t'\simeq \tbar$ and the original cigar
region disappears completely just beyond. This, once again, is the
tunneling transition. We refer to the resulting figure as a `BFKL
duck'.

In the lower plot of figure~\ref{fig:ducks} we show two quantities as
a function of $Y$ (on the same $Y$ scale as the upper plot): the
effective exponent $\om_\mathrm{eff}$, defined as
\begin{equation}
  \om_\mathrm{eff} = \frac{d}{dY} \ln \int dt_0\, dt_1 \,
  e^{-(t_0-t)^2/2-(t_1-t)^2/2} \, G(Y,t_0,t_1)\,,
\end{equation}
and the position of the maximum in $t'$ of $\fbar (Y,Y/2,t,t')$, which
we refer to as $t_{\mathrm{max}}$. At the point where the tunneling
transition occurs there is a jump in the value of $t_\mathrm{\max}$ as
it switches from the perturbative to the non-perturbative region.
Simultaneously there is a rapid change in $\om_\mathrm{eff}$ as it
goes from the perturbative exponent $\simeq \oms$ to the
non-perturbative one $\omp$.

\begin{figure}[htbp]
  \begin{center}
    \epsfig{file=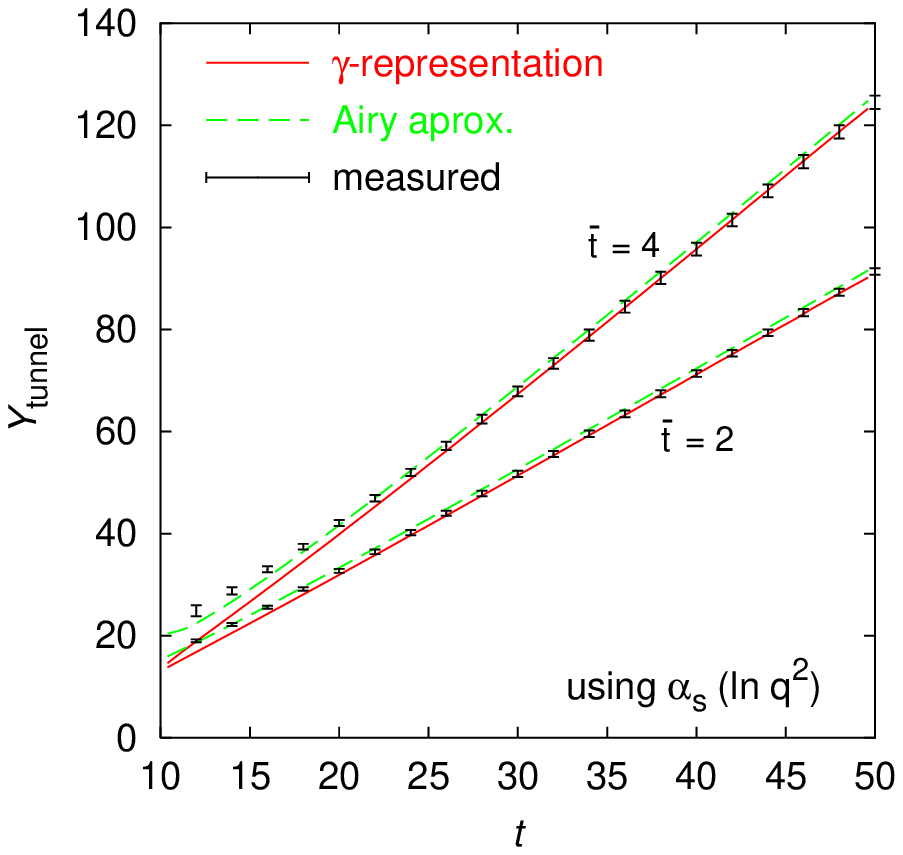,width=0.49\textwidth}
    \epsfig{file=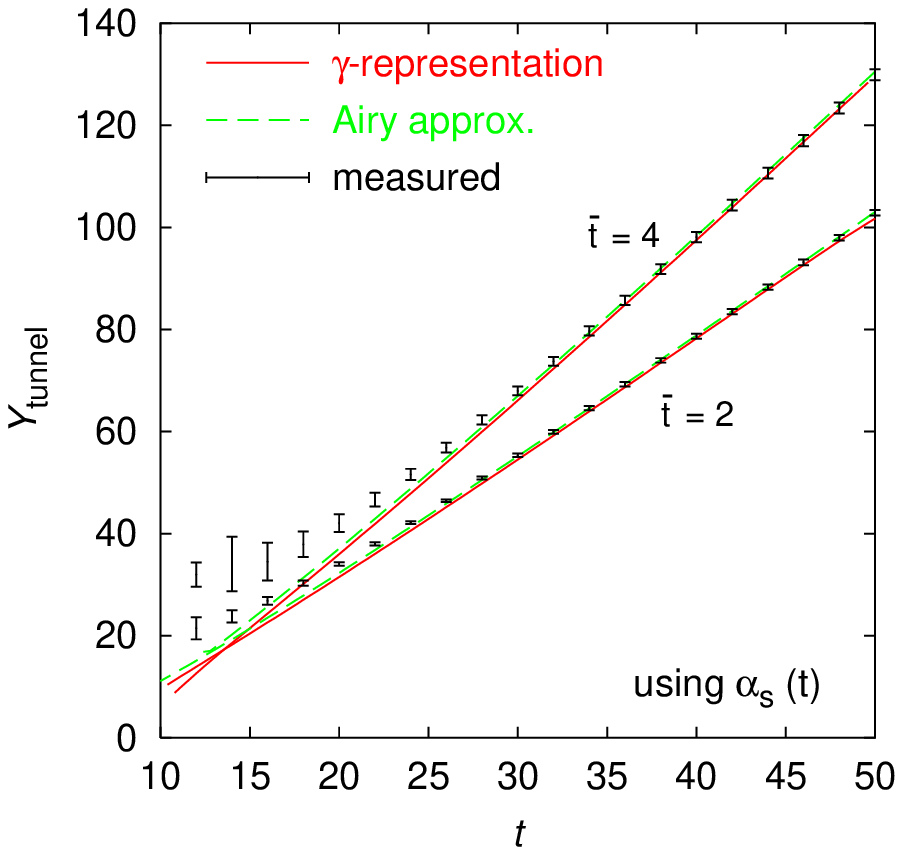,width=0.49\textwidth}
    \caption{The position in $Y$ of the tunneling transition,
      $Y_\mathrm{tunnel}$ as a function of $t$ for two different
      non-perturbative cutoffs $\tbar$ (points).  The points represent
      the numerical `measurements' of $Y_\mathrm{tunnel}$ based on
      the criteria discussed in the text. They are compared to the
      prediction of Eq.~\eqref{eq:YTunnelExact}, dashed lines and the
      calculation which involves the direct numerical evaluation of
      the $\gamma$-representation in the normalisation integrals
      Eq.~(\ref{eq:NormIntegral}), solid lines. Left-hand plot: the
      calculation for $\asb(\ln q^2)$ scale choice, right-hand plot:
      $\asb(\ln k^2)$.}
    \label{fig:ytunnel}
  \end{center}
\end{figure}

We can use the point where the position of the maximum goes from being
above to below $t'=\frac12(t + \tbar)$ as the definition of where
tunneling occurs, $Y_\mathrm{tunnel}$ (in what follows, since we will
no longer be faced with the problem of visualising an initial
$\delta$-function, we will actually consider $f(Y,Y/2,t,t')$ rather
than $\fbar$, though the difference in practice is minimal).
Alternatively one can define the tunneling transition as occurring
when the integrals of $f(Y,Y/2,t,t')$ above and below $\frac12(t +
\tbar)$ are equal,
\begin{equation}
  \int^{(t+\tbar)/2}_{-\infty} dt' f(Y,Y/2,t,t') 
       = \int_{(t+\tbar)/2}^{\infty} dt' f(Y,Y/2,t,t')\,.
\end{equation}
These two definitions for $Y_\mathrm{tunnel}$ have been used to
deduce, from our numerical solutions, the points with errors bars in
fig.~\ref{fig:ytunnel} (the lower error bar corresponds to the first
definition, while the upper error bar corresponds to the second one),
where we plot $Y_\mathrm{tunnel}$ as a function of $t$, for two values
of $\tbar$.  
They are to be compared to the smooth lines, which
correspond to the analytical predictions both in $\asb(t)$ and $\asb(\ln q^2)$ cases.
We have tried  the direct evaluation of the $G_{\pom}$ from Eqs.(\ref{eq:NormSymm}),
(\ref{eq:NormASymm}) by numerically calculating the $\gamma$-representation integrals (solid lines) and the analytical approximation based on the Airy model 
Eq.~\eqref{eq:YTunnelExact} (dashed lines).

One sees a remarkable agreement between the `numerically measured' and
analytical results, which would seem to be strong confirmation of even
the fine details of our picture. We note in particular the roughly
linear dependence of $Y_\mathrm{tunnel}$ on $t$ and the fact that its
slope increases for larger $\tbar$ as a consequence of the smaller
$\omp$ and hence smaller denominator in
eq.~\eqref{eq:YTunnelSimple}.

\section{Conclusions}

In this paper we have presented a detailed investigation of the abrupt
tunneling transition 
from the perturbative regime to the non-perturbative `Pomeron' regime
for the case of leading-log BFKL evolution with a running coupling.

The tunneling transition takes place when $\oms Y \simeq \tbar$, with
$\exp(\tbar)$ the effective non-perturbative scale of the problem (in
units of $\Lambda^2_{QCD}$). This is parametrically much earlier than
estimates of the maximum perturbatively accessible $Y$ value that are
based on diffusion arguments. Furthermore, in contrast to expectations
from diffusion arguments, the transition does not involve evolution at
scales intermediate between those of the hard probe and the
non-perturbative region, as shown in figs.~\ref{fig:cigars} and
\ref{fig:ducks}.

Since the purpose of this paper has been to give an illustration of
the effect of tunneling, we have restricted ourselves to
leading-logarithmic evolution and considered only very extreme scales
for the hard-probe, where the effect is more dramatic because of the
large ratio $\omp/\oms(t)$, or equivalently $t/\tbar$. 

In a more phenomenologically relevant context there remain a number of
other important issues: the separation between tunneling and diffusion
may be less distinct, both because the accessible range of $t$ is
smaller, and because for a given ratio of $t/\tbar$ higher-order
corrections tend to reduce $\omp/\oms(t)$.
Furthermore unitarity and saturation corrections may actually cause
the effective value of $\omp$ to be smaller than $\oms$ which could
eliminate tunneling altogether. Developing a better understanding of
these and related questions remains one of the main challenges of small-$x$
physics.

\section*{Acknowledgements} 

One of us (GPS) wishes to thank the Dipartimento di Fisica,
Universit\`a di Firenze, for hospitality while part of this work was
carried out.

\section*{Appendix A: Pomeron weight with $\boldsymbol{\asb(\ln q^2)}$}

Taking $\asb(\ln q^2)$ instead of $\asb(t)$, as in Eq.(\ref{eq:Bfkl})
means adding a term $-\frac{b}{2} (\chi'_0 + \chi_0^2)$ to the NL
kernel which emerges due to the evolution from scale $t$ to scale $q^2$.
Correspondingly, in the $\omega$ - expansion formulation \cite{CC2,CCS2}
the effective characteristic function becomes
\begin{equation}
\chi_{\om} = \chi_0(\gamma) + \om \frac{1}{\chi_0} \left[-\frac{b}{2}
  (\chi'_0 + \chi_0^2) \right] = \chi_0(\gamma) \left(1 -
  \frac{b\om}{2} \right) - \frac{b\om}{2} \frac{\chi'_0}{\chi_0} \; ,
\end{equation}
so that
\begin{equation}
X_{\om}(\gamma) = \left(1 - \frac{b\om}{2}\right) \left[\log
  \left(\frac{\Gamma(1-\gamma)}{\Gamma(1+\gamma)}\gamma\right)  +  2
  \psi(1) \left(\gamma- \frac{1}{2}\right)  \right] - \frac{b \om }{2
  }\log \frac{\chi_0(\gamma)}{\chi_0(\frac{1}{2})} \; .
\end{equation}
As a consequence, the large $t$-anomalous dimension stays the same
\begin{equation}
b\om t = \chi_{\om}(\gambar) \simeq \frac{1}{\gambar} \; .
\end{equation}
but the saddle point exponent
$\log(\frac{G_{\mathrm{I\!P}}^{(q)}}{N(\omp,\tbar;t)})$ in
(\ref{eq:GPomEstimate}) changes
\begin{multline}
 -t +\frac{2}{b\omp} - \frac{2}{b\omp} \left\{\left[
    \log\left(\gambar\frac{\Gamma(1-\gambar)}{\Gamma(1+\gamma)}\right)
    + 2\psi(1) (\gambar-\frac{1}{2})   \right] \left(1 -
    \frac{b\omp}{2}\right) - \frac{b\omp}{2}\log
  \frac{\chi_m(\gambar)}{\chi_m} \right\} \nonumber \\
 \simeq  -t + \frac{2}{b\omp} \left[ 1 + \psi(1) + \log b\omp t -
  \frac{b\omp}{2} \log \chi_m + \order{\frac{1}{(b\omp t)^2}} \right]
\; .
\end{multline}
It implies that approximately the eigenfunctions $\CF_{\omp}^{(q)}(t)$
get rescaled by the additional constant normalisation factor so that
$\sqrt{\chi_m} \CF_{\omp}^{(q)} \simeq \CF_{\omp}^{(k)} $.



\begin{thebibliography}{99}
\bibitem{LLBFKL} L. N. Lipatov, {\it Sov. J. Nucl. Phys.}  
                  {\bf 23}   (1976) 338;\\  
                  E. A. Kuraev, L. N. Lipatov,  V. S. Fadin,  
                  {\it Sov. Phys. JETP} {\bf 45}  (1977) 199;\\ 
                  I. I. Balitsky,  L. N. Lipatov,  
                  {\it Sov. J. Nucl. Phys.} {\bf 28} (1978) 338.


\bibitem{JK} J.~Kwieci\'nski, {\it Z. Phys. } {\bf C29}  (1985)  561. 

\bibitem{COLKWIE} J.C.~Collins, J.~Kwieci\'nski {\it Nucl.\ Phys.} {\bf B316}  (1989) 307.

\bibitem{Kwiecinski:1999hv}
J.~Kwiecinski, A.~D.~Martin and J.~J.~Outhwaite,
Eur.\ Phys.\ J.\ C {\bf 9} (1999) 611.

\bibitem{LUND} B.~Andersson, G.~Gustafson and  H.~Kharraziha,  
{\it Phys.\ Rev.\ } {\bf D  57} (1998) 5543.



\bibitem{NLLBFKL} 
V.~S.~Fadin, M.~I.~Kotsky, R.~Fiore,  
{\it Phys.\ Lett.\ } {\bf B 359},  (1995) 181;\\ 
%
V.~S.~Fadin, M.~I.~Kotsky, L.~N.~Lipatov, BUDKERINP-96-92,  
{\tt hep-ph/9704267};  \\
%
V.~S.~Fadin, R.~Fiore, A.~Flachi, M.~I.~Kotsky,  
{\it Phys.\ Lett.\ } {\bf B  422},  (1998) 287; \\ 
%
V.~S.~Fadin, L.~N.~Lipatov,  
{\it Phys.\ Lett.\ } {\bf B  429},  (1998) 127;  \\ 
%
%
G.~Camici, M.~Ciafaloni,  
{\it Phys.\ Lett.\ } {\bf B  386},  (1996)341;  
{\it Phys.\ Lett.\ } {\bf B  412},  (1997)  396  
[Erratum-ibid.\ B {\bf 417},  (1997)390  ];  
{\it Phys.\ Lett.\ } {\bf B  430},  (1998) 349.


\bibitem{Salam:1998tj}
G.~P.~Salam,
JHEP {\bf 9807} (1998) 019.

\bibitem{CC2} M.~Ciafaloni, D.~Colferai, Phys. Lett. B {\bf 452} (1999) 372.


\bibitem{CCS2} M.~Ciafaloni, D.~Colferai and G.~P.~Salam,
Phys. Rev. D {\bf 60} (1999) 114036.

\bibitem{ABF} 
G.~Altarelli, R.~D.~Ball and S.~Forte,
Nucl.\ Phys.\ B {\bf 621} (2002) 359;
%
G.~Altarelli, R.~D.~Ball and S.~Forte,
Nucl.\ Phys.\ B {\bf 599} (2001) 383;
%
G.~Altarelli, R.~D.~Ball and S.~Forte,
Nucl.\ Phys.\ B {\bf 575} (2000) 313.

\bibitem{FRSV}  J.~R.~Forshaw, D.~A.~Ross and A.~Sabio Vera,
Phys.\ Lett.\ B {\bf 455} (1999) 273.

\bibitem{BFKLP} 
S.~J.~Brodsky, V.~S.~Fadin, V.~T.~Kim, L.~N.~Lipatov and G.~B.~Pivovarov,
JETP Lett.\  {\bf 70} (1999) 155;
V.~S.~Fadin,
Nucl.\ Phys.\ A {\bf 666} (2000) 155.

\bibitem{SCHMIDT} C.~R.~Schmidt,
Phys.\ Rev.\ D {\bf 60} (1999) 074003.




\bibitem{CCSS2} M.~Ciafaloni, D.~Colferai, G.~P.~Salam and A.~M.~Sta\'sto, in preparation.


\bibitem{Levin:1991xa}
E.~M.~Levin, G.~Marchesini, M.~G.~Ryskin and B.~R.~Webber,
Nucl.\ Phys.\ B {\bf 357} (1991) 167.


\bibitem{Bartels:1993du}
J.~Bartels and H.~Lotter,
Phys.\ Lett.\ B {\bf 309} (1993) 400.

\bibitem{Mueller:1997hm}
A.~H.~Mueller,
Phys.\ Lett.\ B {\bf 396} (1997) 251.

\bibitem{Levin:1995di}
E.~Levin,
Nucl.\ Phys.\ B {\bf 453} (1995) 303;{\it Nucl.~Phys.} {\bf B545}  (1999) 481.  


\bibitem{Kovchegov:1998ae}
Y.~V.~Kovchegov and A.~H.~Mueller,
Phys.\ Lett.\ B {\bf 439} (1998) 428.

\bibitem{Armesto:1998} N.~Armesto, J.~Bartels, M.A.~Braun, {\it Phys.~Lett.} {\bf B442}  (1998) 459. 

\bibitem{Ciafaloni:2001db}
M.~Ciafaloni,  A.~H.~Mueller and  M.~Taiuti,
Nucl.\ Phys.\ B {\bf 616} (2001) 349.


\bibitem{CCSS1} M.~Ciafaloni, D.~Colferai, G.~P.~Salam and
  A.~M.~Sta\'sto, hep-ph/0204282.

\bibitem{Ciafaloni:1999au}
M.~Ciafaloni, D.~Colferai and G.~P.~Salam,
JHEP {\bf 9910} (1999) 017.



\bibitem{CC1}  G.~Camici, M.~Ciafaloni, {\it Phys.~Lett.} {\bf B395}  (1997) 118.

\bibitem{GRADRYZHIK} I.~S.~Gradshtein, I.~M.~Ryzhik, Table of Integrals.

\bibitem{DL}   A. Donnachie, P.V. Landshoff, {\it  Phys. Lett.} {\bf B296}  (1992) 227.



\end{thebibliography}
\end{document}